\documentclass[aps,preprintnumbers,amsmath,nofootinbib,amssymb]{revtex4}

\usepackage{graphics}
\usepackage{epsfig,epsf,psfrag}
\usepackage{times}
\usepackage{float}
\usepackage{color}
\usepackage{amsfonts,amssymb,stmaryrd,latexsym,amsmath}
\usepackage{multirow}
\usepackage{array}
\usepackage{hhline}
\usepackage{booktabs}
\usepackage{enumerate}
\usepackage{slashed}
\usepackage{subfigure}
\usepackage[colorlinks, citecolor=blue,anchorcolor=red,menucolor=red, linkcolor=red,filecolor=red,runcolor=red,urlcolor=blue,frenchlinks=red]{hyperref}

\begin{document}

\bibliographystyle{unsrt}

\title{Threshold effects as the origin of $Z_{cs}(4000)$, $Z_{cs}(4220)$ and $X(4700)$ observed in $B^+\to J/\psi \phi K^+$}

\author{Ying-Hui~Ge$^{1}$}
\author{Xiao-Hai~Liu$^{1}$}~\email{xiaohai.liu@tju.edu.cn}
\author{Hong-Wei Ke$^{1}$}~\email{khw020056@tju.edu.cn}

\affiliation{
    $^1$Center for Joint Quantum Studies and Department of Physics, School of Science, Tianjin University, Tianjin 300350, China
}
\date{\today}

\begin{abstract}

We investigate the $B^+\to J/\psi \phi K^+$ decay via various rescattering diagrams. Without introducing genuine exotic resonances, it is shown that the $Z_{cs}(4000)$, $Z_{cs}(4220)$ and $X(4700)$ reported by the LHCb collaboration can be simulated by the $J/\psi K^{*+}$, $\psi^\prime K^+$ and $\psi^\prime \phi$ threshold cusps, respectively. These cusps are enhanced by some nearby triangle singularities. The $X(4685)$ with $J^P=1^+$ cannot be well simulated by the threshold effects in our model, which implies that it may be a genuine resonance.




\end{abstract}

\maketitle

\section{Introduction}
Very recently, the LHCb collaboration reported the observation of several resonance-like structures in the $B^+\to J/\psi \phi K^+$ decay~\cite{Aaij:2021ivw}. The state $Z_{cs}(4000)^+$ with quark content $c\bar{c} u\bar{s}$ decaying into $J/\psi K^+$ is firstly reported with high significance. Another state $Z_{cs}(4220)^+$ is also reported with significance exceeding five standard deviations. In the $J/\psi\phi$ invariant mass spectrum, four previously reported states $X(4140)$, $X(4274)$, $X(4500)$ and $X(4700)$ are confirmed \cite{Aaij:2016iza,Aaij:2016nsc}. In addition, two new states, $X(4685)$ and $X(4630)$, are firstly reported. What we are interested in this work are the natures of $Z_{cs}(4000)$, $Z_{cs}(4220)$, $X(4700)$ and $X(4685)$.    Their masses, widths and favorable quantum numbers are
\begin{eqnarray}\label{Xstates}
		Z_{cs}(4000):&&\ M=4003\pm 6^{+4}_{-14}\ \mbox{MeV},\ 	\Gamma=131\pm 15\pm 26\ \mbox{MeV},\
	J^{P}=1^{+},  \nonumber \\
	Z_{cs}(4220):&&\ M=4216\pm 24^{+43}_{-30}\ \mbox{MeV},\ 	\Gamma=233\pm 52^{+97}_{-73}\ \mbox{MeV},\
	J^{P}=1^{+},  \nonumber \\
		X(4700):&&\ M=4694\pm 4^{+16}_{-3}\ \mbox{MeV},\ 	\Gamma=87\pm 8^{+16}_{-6}\ \mbox{MeV},\
	J^{P}=0^{+},  \nonumber \\
	X(4685):&&\ M=4684\pm 7^{+13}_{-16}\ \mbox{MeV},\ 	\Gamma=126\pm 15^{+37}_{-41}\ \mbox{MeV},\
	J^{P}=1^{+}.
\end{eqnarray}

Another exotic state $Z_{cs}(3985)$ has ever been reported by BESIII collaboration in 2020~\cite{Ablikim:2020hsk}. It is observed in the $K^+$ recoil-mass spectra in $e^+e^-\to K^+(D_s^- D^{*0}+ D_s^{*-}D^0)$. The mass and width of $Z_{cs}(3985)$ are $3982.5^{+1.8}_{-2.6}\pm 2.1$ MeV and $12.8^{+5.3}_{-4.4}\pm 3.0$ MeV, respectively. Concerning the nature of $Z_{cs}(3985)$, there have been many theoretical interpretations, such as $D_s D^*$ ($D_s^*D$) molecule \cite{Meng:2020ihj,Yang:2020nrt,Du:2020vwb,Chen:2020yvq,Sun:2020hjw,Guo:2020vmu,Yan:2021tcp}, tetraquark state~\cite{Wan:2020oxt,Wang:2020rcx,Wang:2020iqt,Jin:2020yjn}, threshold effects~\cite{Yang:2020nrt,Ikeno:2021ptx}, reflection effects~\cite{Wang:2020kej}, and so on.
From the LHCb observations, one can see that the mass of $Z_{cs}(4000)$ is very close to that of $Z_{cs}(3985)$, but the width of $Z_{cs}(4000)$ is about one order of magnitude larger than that of $Z_{cs}(3985)$. This substantial difference implies that $Z_{cs}(3985)$ and $Z_{cs}(4000)$ may have different origins. The resonance-like peaks observed in the $J/\psi\phi$ invariant mass distributions are also very intriguing, because they may contain both a $c\bar{c}$ pair and and an $s\bar{s}$ pair, which implies that these states may be exotic.

There have been many theoretical interpretations concerning the nature of these exotic hadron candidates, such as molecular states, tetraquark states, or hybrid. Apart from these genuine resonances interpretations, some non-resonance interpretations were also proposed in literature. The threshold effects, such as the threshold cusp and triangle singularity (TS) of the amplitude, can also result in some resonance-like structures in the pertinent invariant mass spectrum, therefore sometimes it is not necessary to introduce a genuine particle to describe a resonance-like peak. The kinematic singularities may simulate genuine resonances, which will bring ambiguities to our understanding about the nature of exotic states.
Before claiming that one resonance-like peak corresponds to one genuine particle, it is also necessary to exclude or confirm these possibilities. We refer to Ref.~\cite{Guo:2019twa} for a recent review about the threshold cusps and TSs in various hadronic reactions.

In this work, we study the $B^+\to J/\psi \phi K^+$ decay by considering several possible rescattering processes, and try to provide a natural explanation for the exotic hadron candidates  $Z_{cs}(4000)$, $Z_{cs}(4220)$, $X(4700)$ and $X(4685)$ reported by LHCb.

\section{Threshold effects and resonance-like structures}
\begin{figure}[htbp]
	\centering
	\includegraphics[width=0.55\hsize]{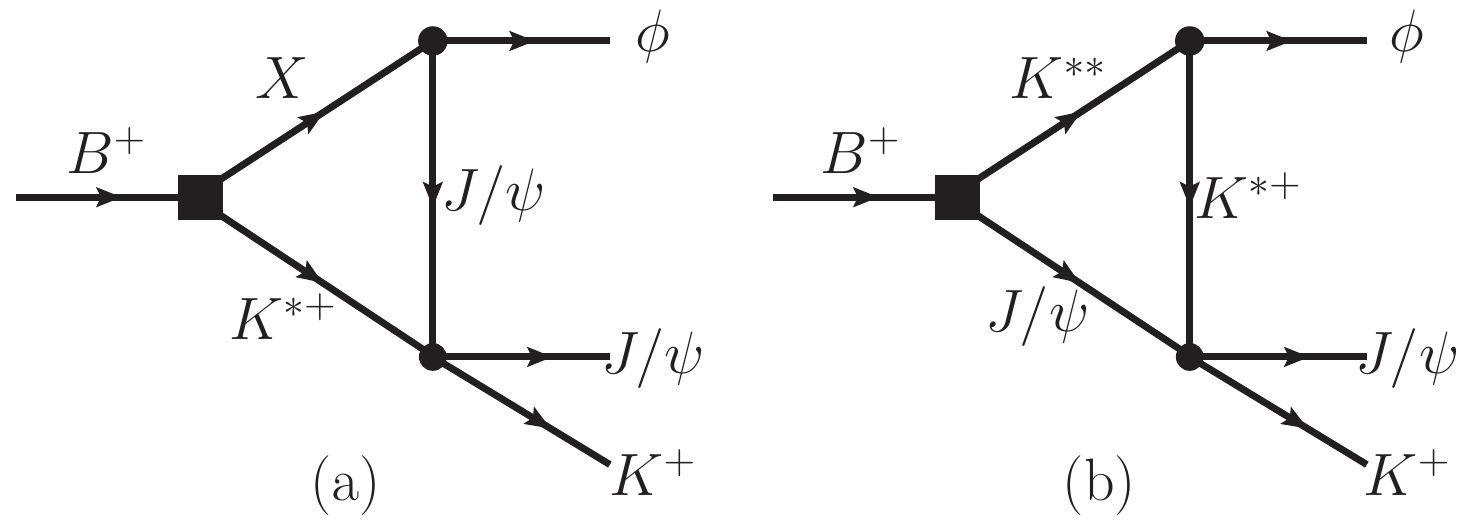}
	\caption{$B^+\to J/\psi \phi K^+$ decay via the (a) $XK^{*}\psi$ and (b) $K^{**}\psi K^*$ rescattering diagrams. Kinematic conventions for the intermediate states are (a) $K^{*+}(q_1)$, $X(q_2)$, $J/\psi (q_3)$ and (b) $J/\psi(q_1)$, $K^{**}(q_2)$, $K^{*+} (q_3)$.}\label{Diagram-Zcs4000}
\end{figure}

\subsection{$Z_{cs}(4000)$}
The bottom meson decaying into a charmonium and a kaon meson is a Cabibbo-favored process. Therefore it is expected that the rescattering processes illustrated in Figs.~\ref{Diagram-Zcs4000}(a) and (b) may play a role in the decay $B^+\to J/\psi \phi K^+$. The intermediate state $X$ in Fig.~\ref{Diagram-Zcs4000}(a) represents any charmonia that can decay into $J/\psi\phi$. From the LHCb experiments one can see that there are many such states. The intermediate state $K^{**}$ in Fig.~\ref{Diagram-Zcs4000}(b) represents a kaon meson that can couple to $\phi K^{(*)+}$. The threshold of $J/\psi K^{*+}$ is about 3989 MeV, which is very close to $M_{Z_{cs}(4000)}$. It is therefore natural to expect that the rescattering process $J/\psi K^{*+}\to J/\psi K^+$ and the resulting threshold cusp may account for the observation of $Z_{cs}(4000)$.

Another intriguing character of the rescattering processes illustrated in Fig.~\ref{Diagram-Zcs4000}(a) is that the $ K^{*+}X$ threshold could be very close to $M_{B^+}$. Therefore a TS of the rescattering amplitude is expected to appear in the vicinity of the physical boundary. The TS may enhance the two-body threshold cusp or itself may generate a resonance-like peak in the $J/\psi K^+$ spectrum.    
The kinematic region where the TS can be present on the physical boundary for various rescattering diagrams is displayed in Table~\ref{KMregion1}, see Ref.~~\cite{Guo:2019twa} for some detailed derivations.
 \begin{table}\caption{TS kinematic region corresponding to the rescattering diagrams in Fig.~\ref{Diagram-Zcs4000}, in unit of MeV.}
 	\begin{center}
 		\begin{tabular}{c|c|c}
 			\hline
 			Diagram & $ M_X / M_{K^{**}} $ & $ M_{J/\psi K^+} $  \\
 			\hline
 			Fig.~\ref{Diagram-Zcs4000}(a)&$M_X$: 4372$\sim$4388 & 3989$\sim$4005   \\
 			\hline
 			Fig.~\ref{Diagram-Zcs4000}(b)&$M_{K^{**}} $: 2068$\sim$2182&  3989$\sim$4099  \\
 			\hline
 		\end{tabular}
 	\end{center}\label{KMregion1}
 \end{table} 
From Table~\ref{KMregion1}, it can be seen that the mass of $X(4274)$ is close to the TS kinematic region for Fig.~\ref{Diagram-Zcs4000}(a). For Fig.~\ref{Diagram-Zcs4000}(b), the mass of $K(1911)$ is relatively close to the TS region.

Considering the $X$ states with $J^P=1^+$, the general invariant amplitude for $B^+\to X K^{*+}$ can be written as:
\begin{eqnarray}\label{BtoXKstar}
	\mathcal{A}(B^+\to X K^{*+}) &=& a\ \epsilon^*(X) \cdot \epsilon^*(K^*) + \frac{b}{(M_B +M_{K^*})^2}\  p_B \cdot \epsilon^*(X) \ p_B \cdot \epsilon^*(K^*) \nonumber\\
	&+& \frac{c}{(M_B +M_{K^*})^2}\ i\varepsilon_{\mu\nu\alpha\beta} p_B^\mu p_{K^*}^\nu \epsilon^{*\alpha}(X) \epsilon^{*\beta}(K^*).
\end{eqnarray}
For $B^+$ decaying into the higher charmonium state $X$ and  $K^{*+}$, the $X$ and  $K^{*+}$ will nearly stay at rest in the rest frame of $B^+$. Therefore in the above formula, only the first term on the right hand side will contribute significantly. As an approximation, we only keep the fist term in the calculation, and set the form factor $a$ as a constant. For the $X$ state with $J^P=1^+$ decaying into $J/\psi \phi $, the amplitude takes the form
\begin{eqnarray}
	\mathcal{A}(X\to J/\psi \phi) = g_{X}  \varepsilon_{\mu\nu\alpha\beta} p_{\phi}^\mu \epsilon^\nu(X) \epsilon^{*\alpha}(J/\psi) \epsilon^{*\beta}(\phi),
\end{eqnarray}
where $g_{X}$ is the coupling constant. To simplify the model, we construct an $S$-wave contact interaction for the scattering $J/\psi K^{*+}\to J/\psi K^+$, which means the quantum numbers of the $J/\psi K^+$ ($J/\psi K^{*+}$) system are $J^P=1^+$. The pertinent amplitude reads
\begin{eqnarray}
	\mathcal{A}(J/\psi K^{*+}\to J/\psi K^+)= g_{\psi K}\ \varepsilon_{\mu\nu\alpha\beta} (p_{J/\psi}^\mu+p_{K}^\mu) \epsilon^\nu(J/\psi) \epsilon^{\alpha}(K^*) \epsilon^{*\beta}(J/\psi).
\end{eqnarray}

The decay amplitude of $B^+\to J/\psi \phi K^+$ via the $XK^{*}\psi$ loop in Fig.~\ref{Diagram-Zcs4000} (a) is given by
\begin{eqnarray}\label{amplitude-Zcs4000}
	&&\mathcal{A}_{B^+\to J/\psi \phi K^+}^{[ XK^{*}\psi]} = -{i} \int \frac{d^4q_1}{(2\pi)^4} \frac{\mathcal{A}(B^+\to X K^{*+})  }{ (q_1^2-M_{{K}^*}^2 +i M_{{K}^*}\Gamma_{{K}^*})  }  \nonumber \\
	&&\times \frac{ \mathcal{A}(X\to J/\psi\phi)\mathcal{A}(J/\psi K^{*+}\to J/\psi K^+) }{ (q_2^2-M_{X}^2 +i M_{X}\Gamma_{X}) (q_3^2-M_{J/\psi}^2) } ,
\end{eqnarray}
where the sum over polarizations of intermediate state is implicit. For the intermediate spin-1 state, the sum over polarization takes the form $\sum_{\mbox{pol}} \epsilon_\mu \epsilon_\nu^*=-g_{\mu\nu}+v_\mu v_\nu$, and we set $v=(1,\boldsymbol{0})$ for a non-relativistic approximation.  The Breit-Wigner type propagators are introduced in Eq.~(\ref{amplitude-Zcs4000}) to account for the width effects of intermediate states.

For Fig.~\ref{Diagram-Zcs4000}(b), considering the $K^{**}$ states with $J^P=1^+$, the invariant amplitude for $B^+\to J/\psi K^{**}$ can be written as
\begin{eqnarray}\label{BtoPsiKstarstar}
	\mathcal{A}(B^+\to J/\psi K^{**}) &=& \tilde{a}\ \epsilon^*(J/\psi) \cdot \epsilon^*(K^{**}) .
\end{eqnarray}
The amplitude for $K^{**}\to \phi K^{*+}$ takes the form
\begin{eqnarray}
	\mathcal{A}(K^{**}\to \phi K^{*+}) = g_{K^{**}}  \varepsilon_{\mu\nu\alpha\beta} p_{\phi}^\mu \epsilon^\nu(K^{**}) \epsilon^{*\alpha}(K^*) \epsilon^{*\beta}(\phi).
\end{eqnarray}
The decay amplitude of $B^+\to J/\psi \phi K^+$ via the $K^{**}\psi K^{*}$ loop in Fig.~\ref{Diagram-Zcs4000} (b) is then given by
\begin{eqnarray}\label{amplitude-Zcs4000-b}
	&&\mathcal{A}_{B^+\to J/\psi \phi K^+}^{[ K^{**}\psi K^{*}]} = -{i} \int \frac{d^4q_1}{(2\pi)^4} \frac{\mathcal{A}(B^+\to J/\psi K^{**})  }{ (q_1^2-M_{J/\psi}^2) }  \nonumber \\
	&&\times \frac{ \mathcal{A}(K^{**}\to \phi K^{*+})\mathcal{A}(J/\psi K^{*+}\to J/\psi K^+) }{ (q_2^2-M_{K^{**}}^2 +i M_{K^{**}}\Gamma_{K^{**}}) (q_3^2-M_{{K}^*}^2 +i M_{{K}^*}\Gamma_{{K}^*})  } .
\end{eqnarray}
\begin{figure}[htbp]
	\centering
	\includegraphics[scale=0.8]{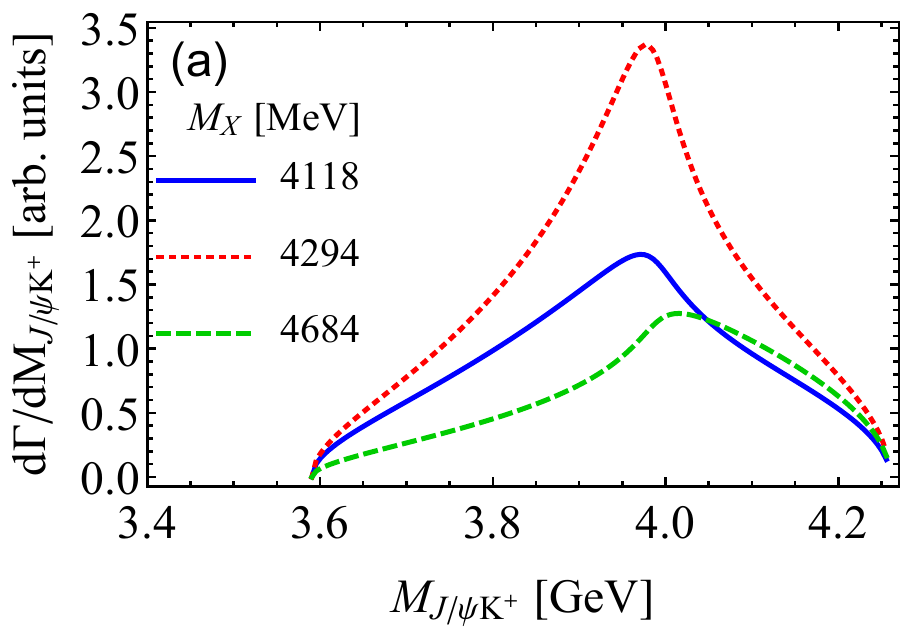} \hspace{0.5cm}
	\includegraphics[scale=0.8]{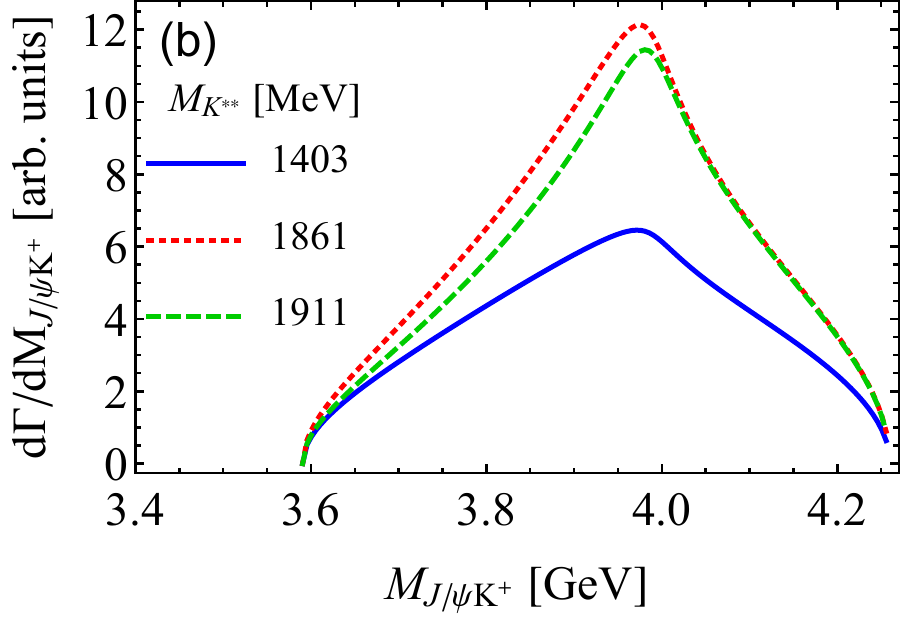}
	\caption{Invariant mass distribution of $J/\psi K^+$ via the rescattering processes in Fig.~\ref{Diagram-Zcs4000}. For (a): The mass and width of intermediate state $X$ are taken to be those of $X(4140)$ (solid line), $X(4274)$ (dotted line), and $X(4685)$ (dashed line) given by LHCb~\cite{Aaij:2021ivw}, separately. For (b): The mass and width of $K^{**}$ are taken to be those of $K_1(1400)$ (solid line), $K(1860)$ (dotted line), and $K(1911)$ (dashed line) given by LHCb~\cite{Aaij:2021ivw}, separately.}\label{lineshape-Zcs4000}
\end{figure}

\begin{figure}[htb]
	\centering
	\includegraphics[scale=0.45]{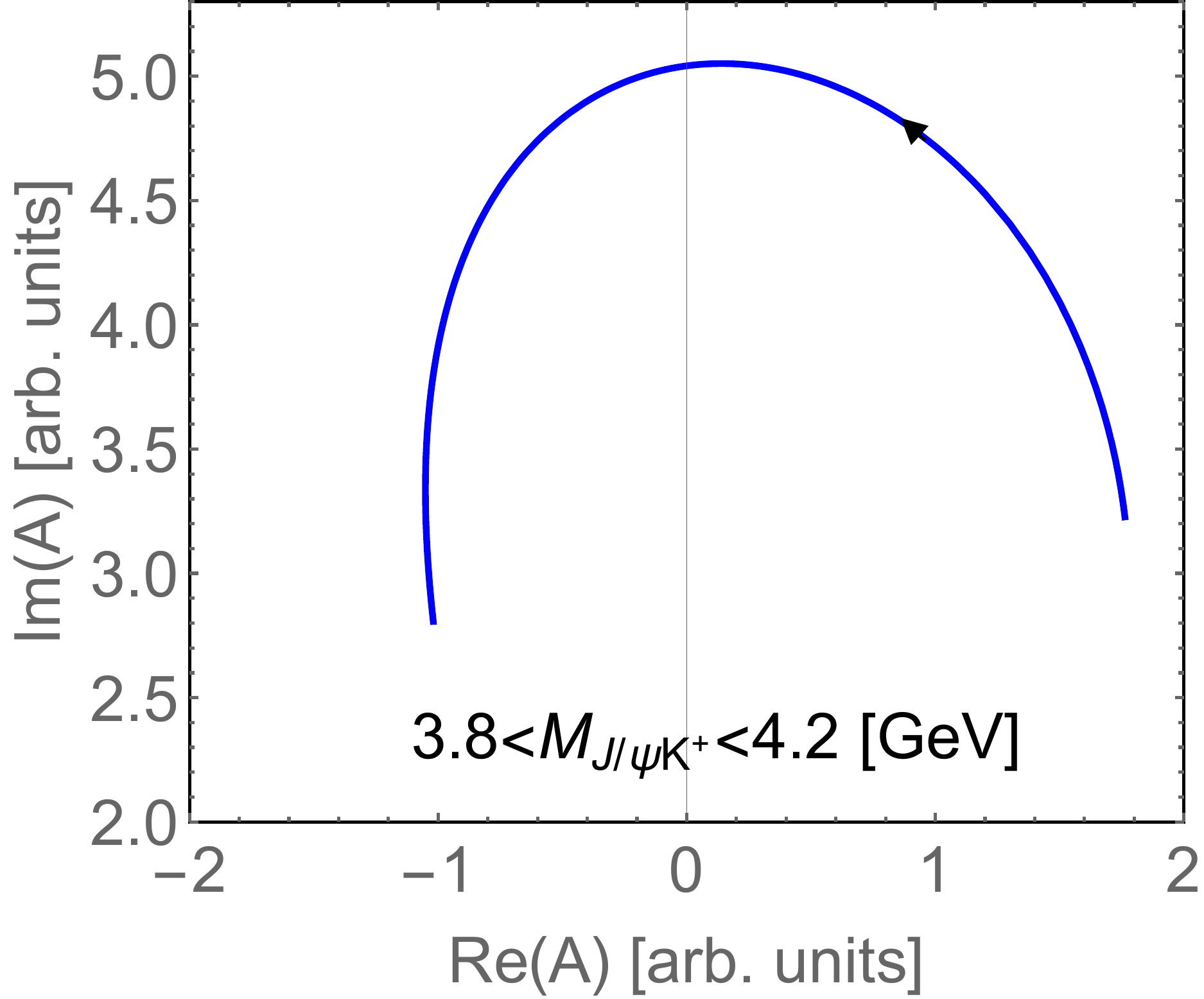}\\
	\caption{Real and imaginary parts of the rescattering amplitude in Eq.~(\ref{amplitude-Zcs4000}). Motion with the increasing invariant mass $M_{J/\psi K^+}$ is counter-clockwise.  }\label{Argand}
\end{figure}

There are many coupling constants in relevant are not well determined yet. Therefore in this work we only focus on the line-shapes of the invariant mass distribution curves. The masses and widths of the known particles involved in the amplitudes are taken from PDG~\cite{Zyla:2020zbs}.
The numerical results of the $J/\psi K^+$ invariant mass distributions via the rescattering processes of Figs.~\ref{Diagram-Zcs4000}(a) and (b) are displayed in Figs.~\ref{lineshape-Zcs4000}(a) and (b), respectively. From Fig.~\ref{lineshape-Zcs4000}(a), one can see that there are bumps around 4 GeV in these distribution curves, which correspond to the $J/\psi K^{*+}$ threshold cusps. These cusps are not quite sharp, because the width of intermediate state $K^{*+}$ of the loop is taken into account in calculating the amplitude, and the cusps are smoothed by the $K^*$ width to some extent. The bumps correspond to the $X(4140)$ and $X(4685)$-loop are broad, while the bump corresponds to the $X(4274)$-loop is relatively narrow and comparable with the width of $Z_{cs}(4000)$ reported by LHCb. This is because the mass of $X(4274)$ is close to the TS kinematic region of the rescattering amplitude, and the threshold cusp is further narrowed by the TS. From this point of view, we can conclude that the threshold effects from the $X(4274)K^{*}\psi$ rescattering loop may simulate the $Z_{cs}(4000)$ structure.

All of the three bumps around $J/\psi K^{*+}$ threshold in Fig.~\ref{lineshape-Zcs4000}(b) are too broad to simulate the $Z_{cs}(4000)$. This is because all of the three $K^{**}$ masses are not close enough to the TS kinematic region of the rescattering amplitude.

The Argand plot corresponds to the rescattering amplitude of $X(4274)$-loop in Eq.~(\ref{amplitude-Zcs4000}) is displayed in Fig.~\ref{Argand}, where the numerator in Eq.~(\ref{amplitude-Zcs4000}) is set to be a constant.  It can be seen that the phase of the amplitude shows a behavior of rapid counter-clockwise change, which is similar to a genuine resonance.

\subsection{$Z_{cs}(4220)$}

\begin{figure}[htbp]
	\centering
	\includegraphics[width=0.55\hsize]{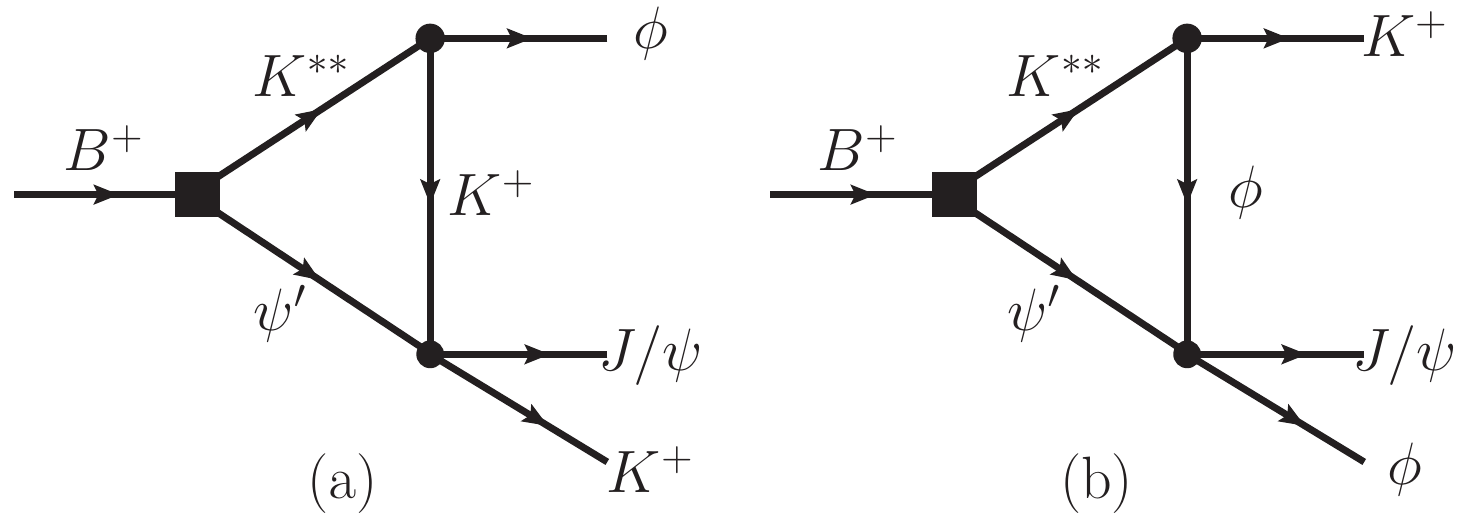}
	\caption{$B^+\to J/\psi \phi K^+$ decay via the (a) $K^{**}\psi^\prime K$ and (b) $K^{**}\psi^\prime \phi$ rescattering diagrams. Kinematic conventions for the intermediate states are (a) $\psi^\prime(q_1)$, $K^{**}(q_2)$, $K^+ (q_3)$ and (b) $\psi^\prime(q_1)$, $K^{**}(q_2)$, $\phi (q_3)$.}\label{Diagram-Zcs4220}
\end{figure}

A relatively broader state $Z_{cs}(4220)$  is also reported by the LHCb. We notice that its mass is close to the $\psi^\prime K^+$ threshold$\sim$4180 MeV. Being similar to the above discussion, one can also expect the rescattering process illustrated in Fig.~\ref{Diagram-Zcs4220}(a) and the resulting $\psi^\prime K^+$ cusp may account for the observation of $Z_{cs}(4220)$. Likewise, the TS kinematic region corresponding to Fig.~\ref{Diagram-Zcs4220}(a) is displayed in Table~\ref{KMregion2}.
For the $K^{**}$ state with spin-1, the invariant amplitude for $B^+\to \psi^\prime K^{**}$ can be written as
\begin{eqnarray}
	\mathcal{A}(B^+\to \psi^\prime K^{**}) &=& a^\prime\ \epsilon^*(\psi^\prime) \cdot \epsilon^*(K^{**}) .
\end{eqnarray}
The amplitudes for $K^{**}(1^+)\to \phi K^{+}$ and  $K^{**}(1^-)\to \phi K^{+}$ are given by
\begin{eqnarray}
	\mathcal{A}(K^{**}\to \phi K^{+}) = g_{A}   \epsilon(K^{**})\cdot \epsilon^{*}(\phi),
\end{eqnarray}
and
\begin{eqnarray}
	\mathcal{A}(K^{**}\to \phi K^{+}) = g_{V}  \varepsilon_{\mu\nu\alpha\beta} p_{K^{**}}^\mu p_{\phi}^\nu \epsilon^\alpha(K^{**})  \epsilon^{*\beta}(\phi),
\end{eqnarray}
respectively.
The amplitude for the near threshold $S$-wave scattering $\psi^\prime K^{+}\to J/\psi K^+$ takes the form
\begin{eqnarray}
	\mathcal{A}(\psi^\prime K^{+}\to J/\psi K^+)= g_{\psi^\prime K}\ \epsilon(\psi^\prime)\cdot  \epsilon(J/\psi).
\end{eqnarray}
The decay amplitude of $B^+\to J/\psi \phi K^+$ via the $K^{**}\psi^\prime K$ loop in Fig.~\ref{Diagram-Zcs4220} (a) is then given by
\begin{eqnarray}\label{}
	&&\mathcal{A}_{B^+\to J/\psi \phi K^+}^{[ K^{**}\psi^\prime K]} = -{i} \int \frac{d^4q_1}{(2\pi)^4} \frac{\mathcal{A}(B^+\to \psi^\prime K^{**})  }{ (q_1^2-M_{\psi^\prime}^2) }  \nonumber \\
	&&\times \frac{ \mathcal{A}(K^{**}\to \phi K^{+})\mathcal{A}(\psi^\prime K^{+}\to J/\psi K^+) }{ (q_2^2-M_{K^{**}}^2 +i M_{K^{**}}\Gamma_{K^{**}}) (q_3^2-M_{K}^2 )  } .
\end{eqnarray}

\begin{table}\caption{TS kinematic region corresponding to the rescattering diagrams in Fig.~\ref{Diagram-Zcs4220}, in unit of MeV.}
	\begin{center}
		\begin{tabular}{c|c|c}
			\hline
			Diagram & $  M_{K^{**}} $ & $ M_{J/\psi K^+}/ M_{J/\psi\phi} $  \\
			\hline
			Fig.~\ref{Diagram-Zcs4220}(a)&1546$\sim$1593 & $M_{J/\psi K^+}$: 4180$\sim$4226  \\
			\hline
			Fig.~\ref{Diagram-Zcs4220}(b)&1572$\sim$1593&  $M_{J/\psi \phi}$: 4706$\sim$4727  \\
			\hline
		\end{tabular}
	\end{center}\label{KMregion2}
\end{table} 

\begin{figure}[htbp]
	\centering
	\includegraphics[scale=0.58]{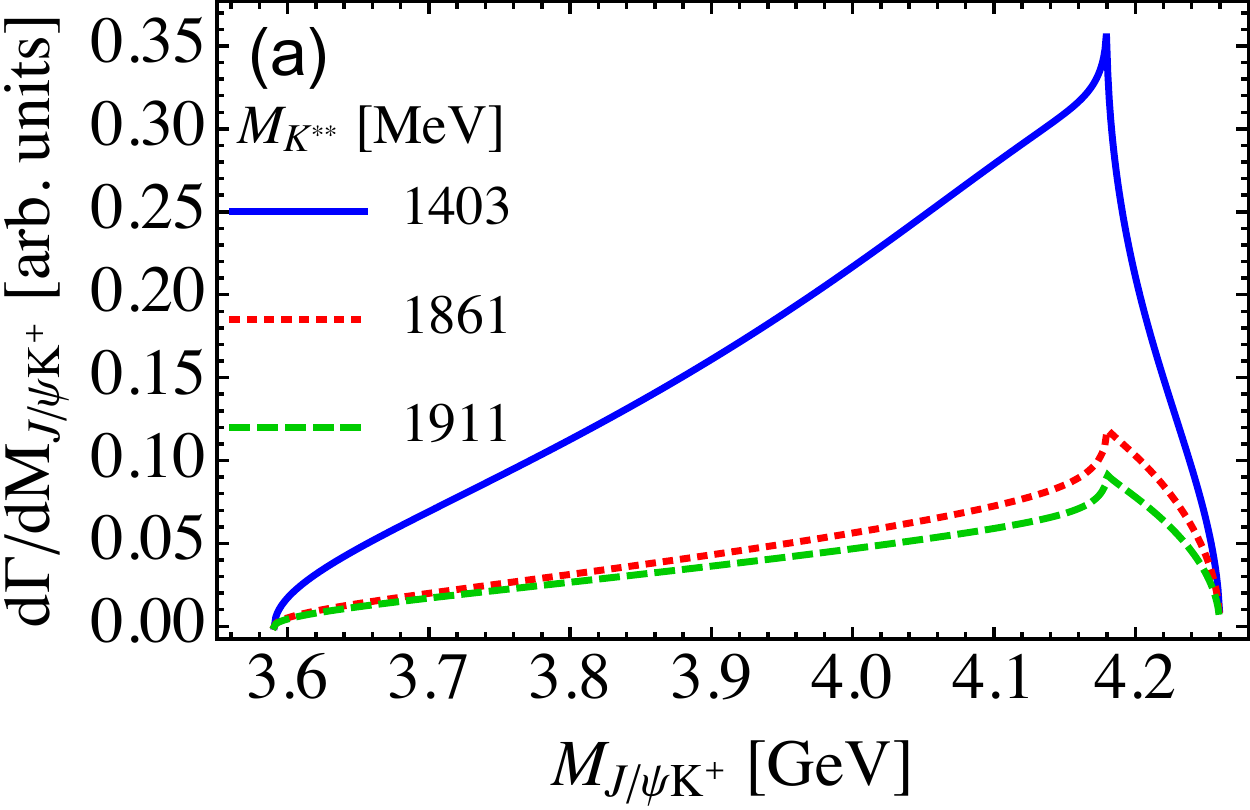} \hspace{0.5cm}
	\includegraphics[scale=0.58]{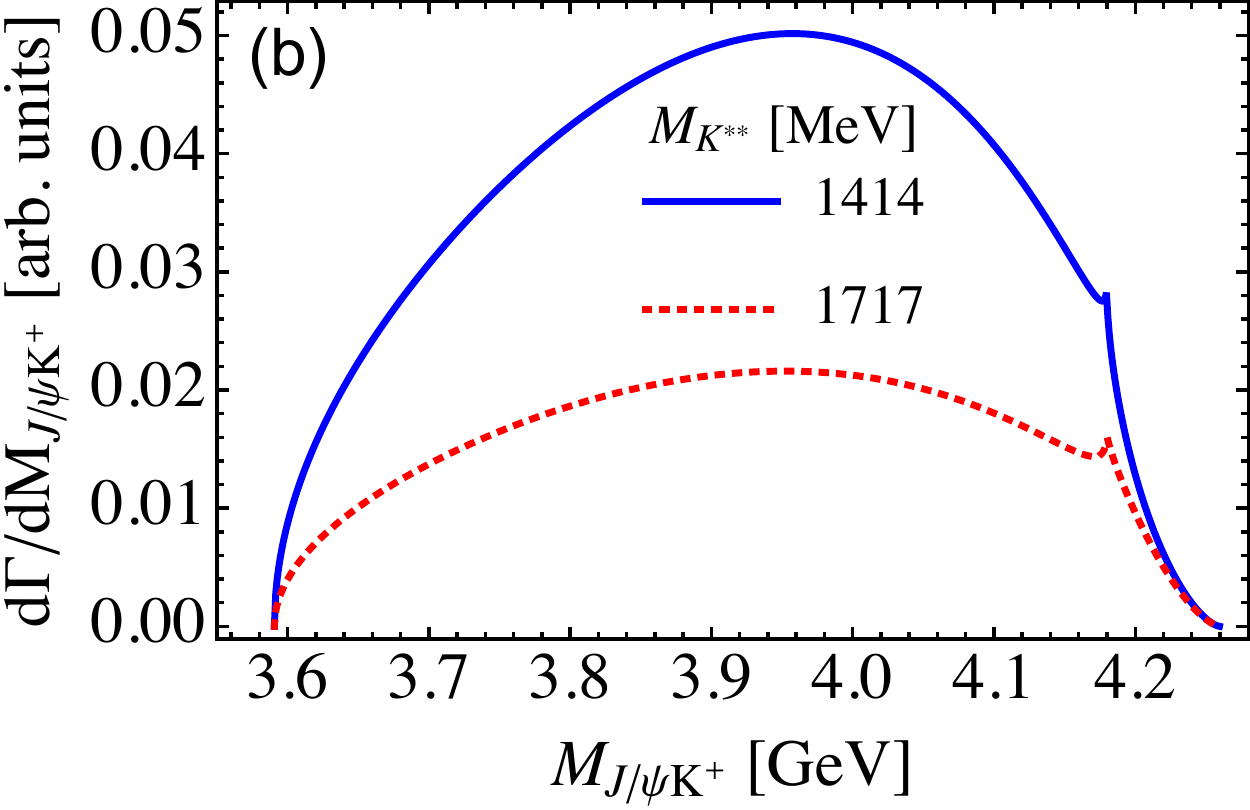}
	\caption{Invariant mass distribution of $J/\psi K^+$ via the rescattering processes in Fig.~\ref{Diagram-Zcs4220}. For (a): The mass and width of $K^{**}$ with $J^P=1^+$ are taken to be those of $K_1(1400)$ (solid line), $K(1860)$ (dotted line), and $K(1911)$ (dashed line) given by LHCb~\cite{Aaij:2021ivw}, separately. For (b): The mass and width of $K^{**}$ with $J^P=1^-$ are taken to be those of $K^*(1410)$ (solid line), and $K^*(1680)$ (dotted line) given by LHCb~\cite{Aaij:2021ivw}, separately.}\label{lineshape-Zcs4220}
\end{figure}

The numerical results of the invariant mass distribution of $J/\psi K^+$ via the rescattering processes in Fig.~\ref{Diagram-Zcs4220} are illustrated in Figs.~\ref{lineshape-Zcs4220}(a) and (b). From Fig.~\ref{lineshape-Zcs4220}(a), one can see that all of three bumps around $\psi^\prime K^+$ threshold are very prominent over the phase space, which may simulate the $Z_{cs}(4220)$ structure. The $\psi^\prime K^+$ threshold cusp corresponds to the $K_1(1400)$-loop is the most prominent one. This is because the mass of $K_1(1400)$ is close to the TS region, as can be seen from Table~\ref*{KMregion2}.

For the Fig.~\ref{lineshape-Zcs4220}(b), one can see that the $\psi^\prime K^+$ cusps correspond to the $K^*(1410)$ and $K^*(1680)$-loop are not very prominent over the phase space. This is because the $K^{**}(1^-)\to \phi K^{+}$ is a $P$-wave decay process, and the rescattering amplitude will be highly suppressed by the small momentum at the edge of the phase space.

\subsection{$X(4700)$ and $X(4685)$}

\begin{figure}[htbp]
	\centering
	\includegraphics[scale=0.8]{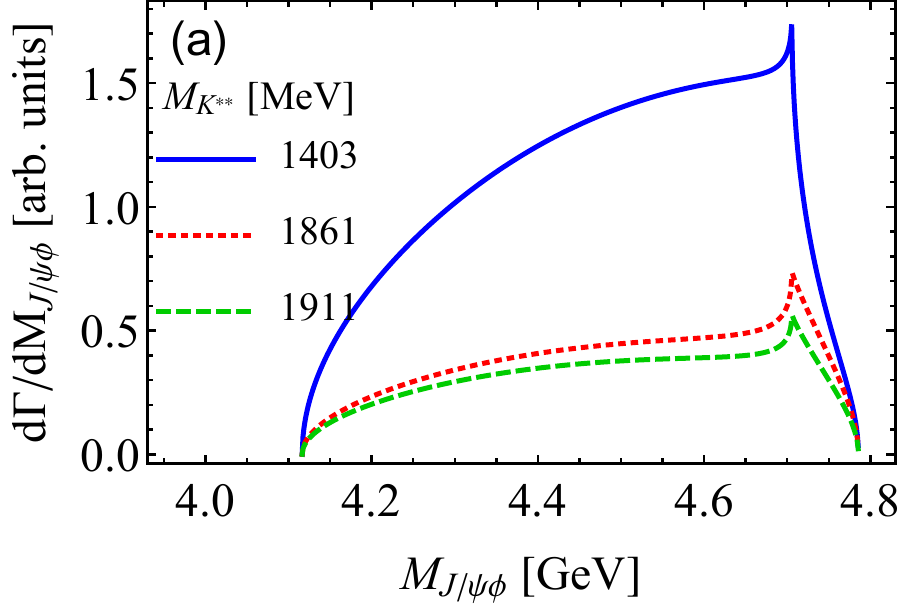} \hspace{0.5cm}
	\includegraphics[scale=0.6]{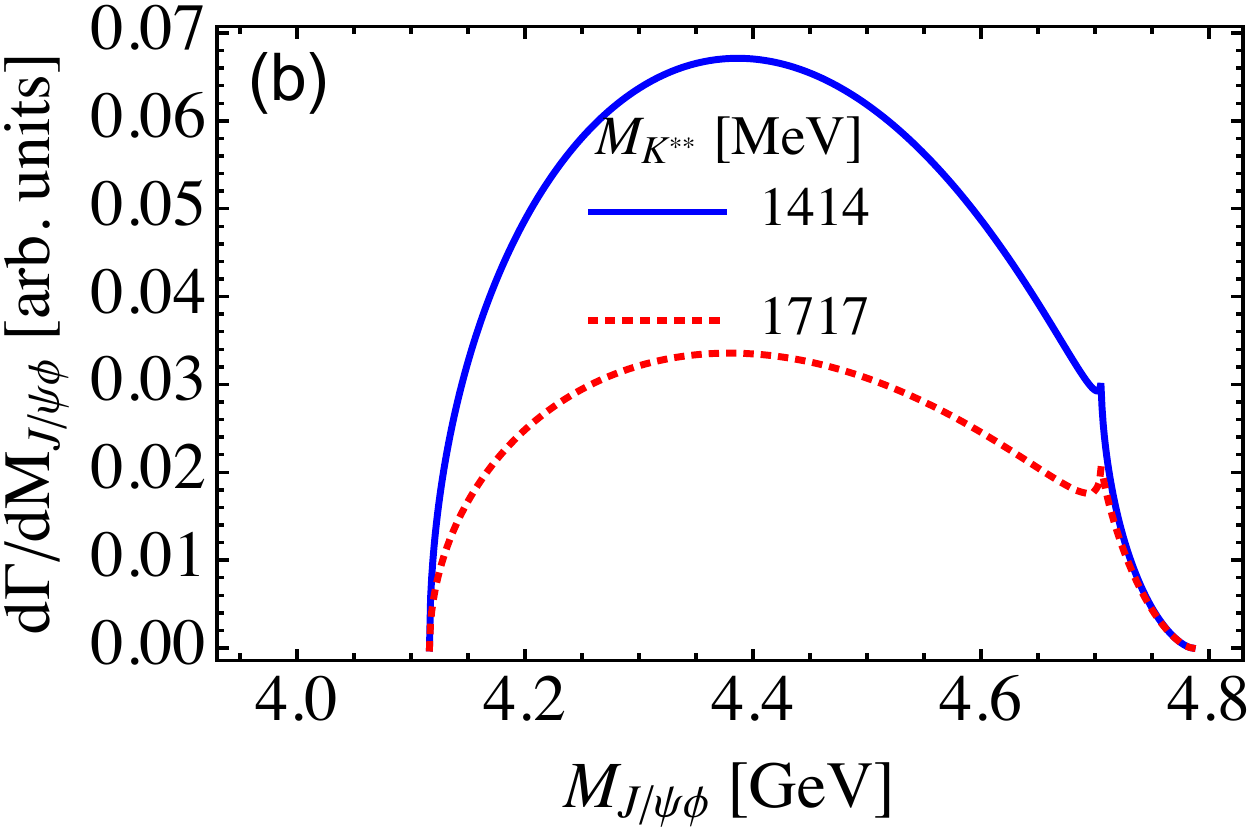}
	\caption{Invariant mass distribution of $J/\psi \phi$ via the rescattering processes in Fig.~\ref{Diagram-Zcs4220}. For (a): The mass and width of $K^{**}$ with $J^P=1^+$ are taken to be those of $K_1(1400)$ (solid line), $K(1860)$ (dotted line), and $K(1911)$ (dashed line) given by LHCb~\cite{Aaij:2021ivw}, separately. For (b): The mass and width of $K^{**}$ with $J^P=1^-$ are taken to be those of $K^*(1410)$ (solid line), and $K^*(1680)$ (dotted line) given by LHCb~\cite{Aaij:2021ivw}, separately.}\label{lineshape-X4700}
\end{figure}

In Ref.~\cite{Liu:2016onn}, we have ever discussed the rescattering diagram illustrated in Fig.~\ref{Diagram-Zcs4220}(b). Since the $\psi^\prime\phi$ threshold (4706 MeV) is very close to the mass of $X(4700)$, and it can be seen from Table~\ref{KMregion2} that $M_{K_1(1400)}$ is close to the TS region, the threshold effect corresponding to the $K_1(1400)\psi^\prime \phi$ rescattering diagram may simulate the $X(4700)$ state, of which the spin-parity is $0^+$. In the updated study of LHCb, another nearby state $X(4685)$ with a broader width and $J^P=1^+$ is also reported. Since the quantum numbers of the $S$-wave $\psi^\prime\phi$ ($J/\psi\phi$) system can be $0^+$, $1^+$ and $2^+$. We try to figure it out whether the threshold effect can also simulate the $X(4685)$.

For the $S$-wave scattering $\psi^\prime\phi \to J/\psi\phi $, the amplitudes for $0^+$ and $1^+$ $J/\psi\phi$ system are given by 
\begin{eqnarray}
	\mathcal{A}(\psi^\prime\phi \to J/\psi\phi )= g_{\psi\phi}\ \epsilon(\psi^\prime) \cdot \epsilon(\phi) \ \epsilon^*(J/\psi) \cdot \epsilon^*(\phi),
\end{eqnarray}
and
\begin{eqnarray}
	\mathcal{A}(\psi^\prime\phi \to J/\psi\phi )= \tilde{g}_{\psi\phi}\ 
	\varepsilon_{\mu\nu\alpha\beta}\varepsilon_{\gamma\delta\lambda\rho} (p_{J/\psi}^\mu+p_{\phi}^\mu)(p_{J/\psi}^\gamma+p_{\phi}^\gamma) g^{\nu\delta}  \epsilon^\alpha(\psi^\prime) \epsilon^{\beta}(\phi) \epsilon^{*\lambda}(J/\psi) \epsilon^{*\rho}(\phi),
\end{eqnarray}
respectively.
In our non-relativistic approximation, if requiring the spin-parity of $J/\psi\phi$ system is $0^+$, only the $1^+$ $K^{**}$ states in Fig.~\ref{Diagram-Zcs4220}(b) will contribute. But if requiring the spin-parity of $J/\psi\phi$ system is $1^+$, only the $1^-$ $K^{**}$ states in Fig.~\ref{Diagram-Zcs4220}(b) will contribute.

The decay amplitude of $B^+\to J/\psi \phi K^+$ via the $K^{**}\psi^\prime \phi$ loop in Fig.~\ref{Diagram-Zcs4220}(b) is then given by
\begin{eqnarray}\label{}
	&&\mathcal{A}_{B^+\to J/\psi \phi K^+}^{[ K^{**}\psi^\prime \phi]} = -{i} \int \frac{d^4q_1}{(2\pi)^4} \frac{\mathcal{A}(B^+\to \psi^\prime K^{**})  }{ (q_1^2-M_{\psi^\prime}^2) }  \nonumber \\
	&&\times \frac{ \mathcal{A}(K^{**}\to \phi K^{+})\mathcal{A}(\psi^\prime \phi\to J/\psi \phi) }{ (q_2^2-M_{K^{**}}^2 +i M_{K^{**}}\Gamma_{K^{**}}) (q_3^2-M_{\phi}^2 )  } .
\end{eqnarray}

The numerical results of the invariant mass distribution of $J/\psi \phi$ via the rescattering processes in Fig.~\ref{Diagram-Zcs4220}(b) are illustrated in Figs.~\ref{lineshape-X4700}(a) and (b). From Fig.~\ref{lineshape-X4700}(a), one can see that the $K^{**}(1^+)\psi^\prime \phi$ loops can generate prominent cusps around 4.7 GeV, which can simulate the $X(4700)$, as discussed in our previous paper~\cite{Liu:2016onn}.

However, for the $K^{**}(1^-)\psi^\prime \phi$ loops shown in Fig.~\ref{lineshape-X4700}(b), the threshold cusps are not prominent over the phase space. This is can be understood. Being similar to the discussion in the last section,
since the $K^{**}(1^-)\to \phi K^{+}$ is a $P$-wave decay process, the rescattering amplitude would be suppressed by a small momentum. Especially around the edge of the phase space, the amplitude would be highly suppressed. If there is no genuine pole around 4685 MeV, the $\psi^\prime\phi$ cusp of the $K^{**}(1^-)\psi^\prime \phi$ rescattering amplitude  itself cannot simulate the $X(4685)$ state with $J^P=1^+$. This implies that the $X(4685)$ observed in the $J/\psi\phi$ distribution could be a genuine resonance. In a very recent paper~\cite{WangZG}, the $X(4685)$ is suggested to be the first radial excited state of the hidden-charm tetraquark $X(4140)$ within the QCD sum rules framework.

\section{Summary}

In summary, we investigate the $B^+\to J/\psi \phi K^+$ decay via various rescattering processes. Without introducing genuine exotic states, it is shown that the $Z_{cs}(4000)$ and $Z_{cs}(4220)$ reported by LHCb can be simulated by the $J/\psi K^{*+}$ and $\psi^\prime K^+$ threshold cusps, respectively. These two cusps are enhanced in the $X(4274)K^{*}\psi$ and $K_1(1400)\psi^\prime K$  rescattering loops, separately. Such phenomena are due to the analytical property of the decaying amplitudes with the TSs located to the vicinity of the physical boundaries. The $X(4700)$ with $J^P=0^+$ can also be simulated by the  $\psi^\prime \phi$ threshold cusp corresponds to the $K^{**}(1^+)\psi^\prime \phi$ loops, as we discussed in a previous paper~\cite{Liu:2016onn}. However, since the contribution of the $K^{**}(1^-)\psi^\prime \phi$ loop is suppressed, the $\psi^\prime \phi$ threshold cusp cannot simulate the $X(4685)$ state with $J^P=1^+$, which implies that the $X(4685)$ could be a genuine resonance.

\begin{acknowledgments}

This work is supported by the National Natural Science Foundation of China (NSFC) under Grant Nos.~11975165 and 12075167. 

\end{acknowledgments}

\end{document}